\documentstyle[prb,aps,preprint]{revtex}
\draft
\begin{document}
\title{Onsager relations and hydrodynamic balance equations}
\author{M.W. Wu, H.L. Cui, and N.J.M. Horing}
\address{Department of Physics and Engineering Physics, Stevens Institute
of Technology, Hoboken, NJ 07030}
\date{\today}
\maketitle
\begin{abstract}
In this paper we clarify the role of heat flux in the hydrodynamic balance
equations, facilitating the formulation of an Onsager relation within the
framework of this theory. Previously thought to be unobtainable from
the present form of the theory
[X.L. Lei, J. Cai, and L.M. Xie, Phys. Rev. B {\bf 38},1529 (1988)], our
verification of the Onsager relation for linear particle and heat flux
currents driven by electric fields and temperature gradients resolves a
puzzling issue of long standing.
Our results show that, for any temperature, when electron density is
sufficiently high, the linear predictions of balance equation theory
exactly satisfy the Onsager relation. The condition of high density is
consonant with the requirement of strong electron-electron interactions for
the validity of the Lei-Ting balance equations. Our results
support the validity of this theory for a weakly nonuniform system.
We also discuss a possible method of extending this theory to
a system further removed from thermal equilibrium.
\end{abstract}


\section{Introduction}

The balance equation transport theory of Lei and
Ting\cite{leiting,leihoring} was originally developed to treat high-field
electrical condition in homogeneous semiconductors, and has achieved much
success in hot-electron semiconductor transport problems. This theory is
based on a separation of the center of mass of the system from the relative
motion of electrons in the presence of a uniform electric field. The center
of mass is treated as a classical particle, whereas the relative electron
system, which is composed
of a large number of interacting particles, is treated fully
quantum-mechanically. The theory has been successfully applied to a variety
of transport problems, and the results obtained have exhibited good agreement
with experiments.\cite{hirakawa} This theory was subsequently
generalized to deal with weakly nonuniform, inhomogeneous systems by Lei
{\em et al.}.\cite{lei} The resulting hydrodynamic balance equations obtained
by them consist of the following
three equations: (a) continuity equation; (b) momentum balance equation;
and (c) energy balance equation.

The form of these hydrodynamic balance equations
appears very similar to their classical counterparts, generally called
hydrodynamic models.\cite{1,2,3,4,5,6,7,8,9,10,11,12} However, in actual
fact, they are quite different. The latter is derived from the Boltzmann
transport equation, as the first three moments of that equation. Very
recently, the fourth moment was discussed by Anile {\em et
al.}\cite{anile1,anile2,anile3}, in an attempt to include the equation
describing heat
flux. Although, in principle, a complete determination of Boltzmann
equation is equivalent to the determination of all the moments, it is not
practical to solve the infinite hierarchy of coupled equations governing the
various moments. The hydrodynamic approach is
based on truncation of this hierarchy after the second order moment, and
simplification of the remaining equations. However these three
moment equations by themselves do not form a closed system, requiring input
of information about scattering, generally supplied in the form
of momentum and energy relaxation times. Nevertheless, to accurately
evaluate the relaxation times requires a predetermination of the distribution
function, which is precisely the task that the hydrodynamic models strive to
avoid. This difficulty is circumvented by one of the following ways. One
approach is to calculate the relaxation times by Monte Carlo simulations.
Another employs empirical forms of relaxation times. The third is to
postulate a distribution function with unknown parameters, and use the
hydrodynamic equations to solve
for these parameters. One of the most popular parameterized distribution
functions is the drifted Maxwell distribution, which depends on two
unknown parameters, the electron drift velocity and the electron
temperature. The hydrodynamic balance equation approach employs a drifted
local equilibrium description similar to the latter. In this
it employs unknown parameters including the local
electron temperature $T_e({\bf R})$, local electron drift velocity ${\bf v}
({\bf R})$ and local chemical potential $\mu({\bf R})$. The distinctive
features of balance equation theory rest with the ansatz of such local
equilibrium parameters in an appropriately chosen initial density matrix,
which is treated quantum mechanically, describing the dynamics of
the many-body system of electrons, impurities and phonons. Of course,
these unknown parameters are also to be determined from
the resulting balance equations. It is now believed that the specific
quasi-equilibrium form of the initial density matrix chosen in balance
equation theory is specifically suited to the condition of strong
electron-electron interactions, since it requires rapid
thermalization about the drifted transport state.\cite{chen1,chen2}
A salient feature of this new hydrodynamic approach is that it includes a
microscopic description of scattering in the form of
a frictional force function due to electron-impurity and electron-phonon
scattering, as well as an electron energy loss rate function due to
electron-phonon
interaction.  These functions are calculated within the model itself, as
functions of carrier drift velocity and carrier temperature, along with
the carrier density, which are themselves determined self-consistently
within the same model. These hydrodynamic balance equations have recently
been applied to device simulations by Cai {\em et al.}.\cite{cai1,cai2,cai3}

A hitherto unresolved question, unanswered since the development of
hydrodynamic balance equations, concerns the capability of this theory
to lead to the correct form of Onsager relations\cite{onsager,mahan}
and/or how to determine Onsager relations within the framework of this
theory. There is even some misunderstanding that the energy flux
predicted by this theory is zero. The purpose of this paper is to clarify
the role of heat flux in this theory, and to also show how to generate
Onsager relations within the framework of this theory. We have
closely checked the Onsager relation predicted by this theory and find, that
for any temperature, when electron density is sufficiently high, the balance
equation theory satisfies Onsager relations exactly. The condition of high
density is consonant with the requirement that Lei-Ting balance equations
hold only for strong electron-electron interactions.
Furthermore, our results support the validity of this
theory in weakly nonuniform systems. To our knowledge, this is the first
set of hydrodynamic equations which obeys Onsager relation exactly.
Anile {\em et al.} showed very recently,\cite{anile3} by Monte Carlo
simulation that the Onsager relation fails in the traditional hydrodynamic
models.

This paper is organized as follows: In Sec.\ II we review the derivation of
the hydrodynamic balance equations. This is not insignificant because we
explicitly exhibit the role of the energy flux in this theory. Moreover, we
also formulate the hydrodynamic force and energy balance equations in
somewhat different forms than those of Lei {\em et al.},\cite{lei} which
clarifies the meaning of every term. Then, in Sec.\ III we derive the
Onsager relation for linear particle and heat flux currents driven by
electric field and temperature gradient, and check it
closely. We present our conclusions and discussions in Sec.\ IV.

\section{Hydrodynamic balance equations}

The starting point of hydrodynamic balance equation theory consists of the
following fluid-element-composed electron Hamiltonian
\begin{equation}
H=\int d{\bf R}\ [H_e({\bf R})+H_I({\bf R})]\;.
\end{equation}
Here,
\begin{equation}
\label{he}
H_e({\bf R})=\sum_i\left[\frac{{\bf p}_i^2}{2m}+\frac{1}{2}\sum_{i\not= j}
\frac{e^2}{|{\bf r}_i-{\bf r}_j|}\right]\delta({\bf r}_i-{\bf R})
\end{equation}
denotes the kinetic energy and Coulomb interaction energy of electrons
within a fluid cell around ${\bf R}$. Macroscopically this cell is small
over which all the expectations of physical quantities change little,
whereas microscopically it is large enough that a great number of particles
are within it. ${\bf p}_i$ and ${\bf r}_i$ are the momentum and coordinate
of the $i$-th electron.
\begin{equation}
\label{hi}
H_I({\bf R})=\sum_i[e\phi ({\bf r}_i)+\Phi ({\bf r}_i)]\delta ({\bf r}_i
-{\bf R})
\end{equation}
is the interaction Hamiltonian in which $\phi ({\bf r})$ denotes the
potential of the external electric field ${\bf E}$, hence ${\bf E}=-\nabla
\phi ({\bf r})$, and $\Phi ({\bf r})=\sum_a u({\bf r}-{\bf R}_a)+
\sum_\ell {\bf u}_\ell\cdot\nabla v_\ell ({\bf r}-{\bf R}_\ell)$ represents
the scattering potential due to randomly distributed (${\bf R}_a$)
impurities and lattice vibrations (${\bf R}_\ell$ stands for the lattice
sites). The number density of electrons in the cell around ${\bf R}$ may be
written as
\begin{equation}
N({\bf R})=\sum_i\delta({\bf r}_i-{\bf R})\;.
\end{equation}
Similarly the ${\bf R}$-dependent momentum density is given by
\begin{equation}
\label{p}
{\bf P}({\bf R})=\sum_i{\bf p}_i\delta({\bf r}_i-{\bf R})\;.
\end{equation}
Letting ${\bf v}({\bf R})$ be the average electron velocity in the
fluid cell about ${\bf R}$, which is a parameter to be determined
self-consistently from the resulting balance equations, one can write the
statistical average of the momentum density as
\begin{equation}
\langle{\bf P}({\bf R})\rangle=mn({\bf R}){\bf v}({\bf R})\;,
\end{equation}
with $n({\bf r})=\langle N({\bf R})\rangle$, the statistical average of the
electron number density. Introducing relative electron variables
\begin{equation}
\label{pprp}
{\bf p}_i^\prime={\bf p}_i-m{\bf v}({\bf R})\hspace{0.5cm},\hspace{0.5cm}
{\bf r}_i^\prime={\bf r}_i-{\bf R}\;,
\end{equation}
which represent the momentum and coordinate of the $i$-th electron relative
to the center of mass of the fluid cell around ${\bf R}$, we can write the
statistical average of $H_e({\bf R})$ as
\begin{equation}
\label{heav}
\langle H_e({\bf R})\rangle=u({\bf R})+\frac{1}{2}mn({\bf R})v^2({\bf R})\;,
\end{equation}
with
\begin{equation}
u({\bf R})=\langle\sum_i\frac{{\bf p}_i^{\prime 2}}{2m}
\delta({\bf r}_i^\prime)\rangle
\end{equation}
denoting the average kinetic energy of the relative electron in cell
${\bf R}$. It is noted that in deriving Eq.\ (\ref{heav}) we have treated
electron-electron Coulomb interaction effect in the spirit of Landau
fermi-liquid theory, which is appropriate for electrons in semiconductors
and metals, {\em ie.}, it leads to a self-energy correction in the
single electron energy, and also renormalizes the bare phonon frequency,
jointly with the bare electron-phonon interaction vertex, and also the
electron-impurity interaction vertex.\cite{scalapino,mahan,callaway}
We assume that these renormalized corrections are already included in the
corresponding quantities. The use of the Hamiltonian above is well
established and similar to those discussed in the book of
Zubarev.\cite{zubarev}

Considering the rate of change of particle number density, $\dot
N({\bf R})=-i[N({\bf R}),H]$, and performing the statistical
average, the equation of continuity follows as
\begin{equation}
\label{eq1}
\frac{\partial n}{\partial t}+\nabla\cdot(n{\bf v})=0\;,
\end{equation}
where we have used the relation
\begin{equation}
\label{rdot}
\dot{\bf r}_i=-i[{\bf r}_i,H]={\bf p}_i/m\;.
\end{equation}
The particle flux density operator ${\bf J}({\bf R})$ can be derived from
the momentum density operator Eq.\ (\ref{p}) as
\begin{equation}
{\bf J}({\bf R})=\frac{1}{m}{\bf P}({\bf R})=\sum_i\frac{{\bf p}_i}{m}
\delta({\bf r}_i-{\bf R})\;,
\end{equation}
and the rate of change of ${\bf J}({\bf R})$ can be written as
\begin{equation}
\label{jdot}
\dot{\bf J}({\bf R})=-i[{\bf J}({\bf R}),H]=\sum_i\frac{1}{m}(e{\bf E}
+{\bf F}_i)\delta({\bf r}_i-{\bf R})-\nabla_{\bf R}\cdot\sum_i
\frac{{\bf p}_i}{m}\frac{{\bf p}_i}{m}\delta({\bf r}_i-{\bf R})\;.
\end{equation}
Here, we have used the relation
\begin{equation}
\label{pdot}
\dot{\bf p}_i=-i[{\bf p}_i,H]=(e{\bf E}-\nabla \Phi({\bf r}_i))/m\equiv
(e{\bf E}+{\bf F}_i)/m\;,
\end{equation}
with ${\bf F}_i$ representing the force operator of the $i$th-electron.
Transforming to the relative coordinate variables, Eq.\ (\ref{pprp}),
and performing the statistical
average of Eq.\ (\ref{jdot}), we have
\begin{equation}
\label{jdotav}
\frac{\partial}{\partial t}\langle {\bf J}({\bf R})\rangle+
\nabla\cdot(\langle{\bf J}({\bf R})\rangle{\bf v})
=-\nabla\cdot\left\langle\sum_i\frac{{\bf p}_i^\prime}{m}
\frac{{\bf p}_i^\prime}{m}\delta({\bf r}_i^\prime)\right\rangle+\frac{e
n({\bf R}){\bf E}}{m}+\frac{{\bf f}({\bf R})}{m}\;,
\end{equation}
where
\begin{equation}
\label{jav}
\langle {\bf J}({\bf R})\rangle=n({\bf R}){\bf v}({\bf R})\;,
\end{equation} and
\begin{equation}
{\bf f}({\bf R})=-\langle\sum_i\nabla\Phi({\bf r}_i^\prime+{\bf R})
\delta({\bf r}_i^\prime)\rangle
\end{equation}
is the frictional force experienced by the fluid cell due to impurity and
phonon scattering. Since
\begin{equation}
\left\langle \frac{{\bf p}_i^\prime}{m}\frac{{\bf p}_i^\prime}{m}
\delta({\bf r}_i^\prime)\right\rangle=\frac{2}{3m}\left\langle
\sum_i\frac{{\bf p}_i^{\prime 2}}{2m}\delta({\bf r}_i^\prime)\right
\rangle {\cal I}=\frac{2}{3m}u({\bf R}){\cal I}\;,
\end{equation}
with ${\cal I}$ as the unit tensor, one follows that
\begin{equation}
\label{eq2}
\frac{\partial}{\partial t}\langle {\bf J}({\bf R})\rangle+\nabla\cdot(
\langle {\bf J}({\bf R})\rangle{\bf v})=-\frac{2}{3m}\nabla u({\bf R})
+\frac{{\bf f}({\bf R})}{m}\;.
\end{equation}
This equation can be proved directly to be just the original Euler-type
momentum balance equation obtained by Lei {\em et al.}:\cite{lei}
\begin{equation}
\label{eq2lei}
\frac{\partial {\bf v}}{\partial t}+{\bf v}\cdot\nabla {\bf v}=-\frac{2}{3}
\frac{\nabla u}{mn}+\frac{e}{m}{\bf E}+\frac{\bf f}{mn}\;,
\end{equation}
if one takes Eq.\ (\ref{eq1}) into account.

Similarly one can derive the energy balance equation by averaging the
Heisenberg equation of motion $\dot H_e({\bf R})=-i[H_e({\bf R}),H]$, which
should be combined with the time derivative of Eq.\ (\ref{heav}), and
yields
\begin{eqnarray}
\frac{\partial u}{\partial t}+\nabla\cdot\langle{\bf J}_H\rangle&=&
\frac{2}{3}{\bf v}\cdot\nabla u+\frac{1}{2}mv^2\nabla\cdot(n{\bf v})
+\frac{1}{2}mn{\bf v}\cdot\nabla v^2\nonumber\\
\label{eq3}
&&-w-{\bf v}\cdot{\bf f}\;.
\end{eqnarray}
Here
\begin{equation}
w({\bf R})=\frac{1}{2}\langle\sum_i\frac{{\bf p}_i^\prime}{m}\cdot
\nabla\Phi({\bf
r}_i^\prime+{\bf R})\delta({\bf r}_i^\prime)\rangle+
\frac{1}{2}\langle\sum_i \nabla\Phi({\bf r}_i^\prime+{\bf R})\cdot
\frac{{\bf p}_i^\prime}{m}\delta({\bf r}_i^\prime)\rangle-{\bf v}({\bf R})
\cdot{\bf f}({\bf R})
\end{equation}
is the energy transfer rate from electron system to phonon system, and
\begin{equation}
\label{jhoper}
{\bf J}_H({\bf R})=\sum_i\frac{{\bf p}_i^2}{2m}\frac{{\bf p}_i}{m}\delta
({\bf r}_i-{\bf R})
\end{equation}
is the energy flux operator, whose statistical average being
\begin{equation}
\label{jhl}
\langle {\bf J}_H({\bf R})\rangle=\frac{5}{3}u({\bf R}){\bf v}({\bf R})
+\frac{1}{2}mn({\bf R})v^2({\bf R}){\bf v}({\bf R})\;.
\end{equation}
This is just the energy flux predicted by balance equation theory.
Taking this equation into account, one can easily recover the original form
of the energy balance equation of Ref.\ \onlinecite{lei} by substituting
Eq.\ (\ref{jhl}) into Eq.\ (\ref{eq3}):
\begin{equation}
\label{eq3lei}
\frac{\partial u}{\partial t}+{\bf v}\cdot\nabla u=-\frac{5}
{3}u(\nabla\cdot{\bf v})-w-{\bf v}\cdot{\bf f}\;.
\end{equation}

The resistive force ${\bf f}$, the energy transfer rate $w$, together with
the local kinetic energy $u$ and the local density $n$ are calculated
within the framework of balance equation theory\cite{leiting}, which
requires knowledge of the density matrix $\hat{\rho}$. This density matrix
can be determined by solving the Liouville equation, $i\partial\hat{\rho}/
\partial t=[H,\hat{\rho}]$, with an appropriate initial
condition. In the balance
equation theory, the electron-impurity and electron-phonon couplings are
turned on from $t=0$, together with the electric field ${\bf E}$. Meanwhile
in the present model the interactions between different fluid cells are
included approximately in the local potential with a mean field treatment.
Therefore different cells are dynamically independent, and thus evolve
separately from their own initial state. Thus, the ${\bf R}$-dependent
initial density matrix is chosen such that the relative electron system in
the fluid cell is in a local quasi-thermal equilibrium state at electron
temperature $T_e({\bf R})$ and chemical potential $\mu({\bf R})$,
which are parameters to be
determined self-consistently from the resulting hydrodynamic balance
equations; whereas the phonon system is assumed in thermal equilibrium:
\begin{equation}
\label{rho0}
{\hat \rho}_0=\frac{1}{Z}\exp\{-\sum_{\bf R}[H_e({\bf R})-{\bf v}({\bf R})
\cdot{\bf P}({\bf R})-\mu]/T_e({\bf R})\}\exp(-H_{ph}/T)
\end{equation}
with $H_{ph}$ and $T$ being the phonon Hamiltonian and temperature. It
follows that the resistive force and the energy transfer rate are given by
\begin{eqnarray}
{\bf f}({\bf R})&=&{\bf f}(n({\bf R}),T_e({\bf R}),{\bf v}({\bf R}))=
n_i\sum_{\bf q}{\bf q}|u({\bf q})|^2\Pi_2({\bf q},{\bf q}\cdot{\bf v}
({\bf R}))\nonumber\\
&&\mbox{}-2\sum_{{\bf q}\lambda}{\bf q}|M({\bf q},\lambda)|^2\Pi_2
({\bf q},\Omega_{{\bf q}\lambda}-{\bf q}\cdot{\bf v}_d)\left[n(\frac{\Omega
_{{\bf q}\lambda}}{T})-n(\frac{\Omega_{{\bf q}\lambda}-{\bf q}\cdot{\bf v}
({\bf R})}{T_e({\bf R})})\right]\;,\\
w({\bf R})&=&w(n({\bf R}),T_e({\bf R}),{\bf v}({\bf R}))\nonumber\\
&=&2\sum_{{\bf q}\lambda}\Omega_{{\bf q}\lambda}
|M({\bf q},\lambda)|^2\Pi_2
({\bf q},\Omega_{{\bf q}\lambda}-{\bf q}\cdot{\bf v}_d)\left[n(\frac{\Omega
_{{\bf q}\lambda}}{T})-n(\frac{\Omega_{{\bf q}\lambda}-{\bf q}\cdot{\bf v}
({\bf R})}{T_e({\bf R})})\right]\;,
\end{eqnarray}
with $n(x)=(e^x-1)^{-1}$ being Bose distribution function; $n_i$, impurity
density; $\Omega_{{\bf q}\lambda}$, the phonon frequency of wave
${\bf q}$ and mode $\lambda$; $u({\bf q})$, the electron-impurity
interaction potential, and $M({\bf q},\lambda)$, the electron-phonon
correction matrix element. $\Pi_2({\bf q},\lambda)$ denotes the imaginary
part of electron density-density correction function. Note that
${\bf f}$ and $w$ depend on ${\bf R}$ through the quantities $n({\bf R})$,
$T_e({\bf R})$ and $v({\bf R})$.
The average local kinetic energy density of the relative electrons is
\begin{equation}
\label{u}
u=2\sum_{\bf k}\varepsilon_{\bf k}f[(\varepsilon_{\bf k}-\mu)/T_e]\;,
\end{equation}
and the local chemical potential $\mu({\bf R})$ is related to the local
density $n({\bf R})$ of electrons through the relation
\begin{equation}
\label{n}
n=2\sum_{\bf k}f[(\varepsilon_{\bf k}-\mu)/T_e]\;,
\end{equation}
with $\varepsilon_{\bf k}=k^2/2m$ and $f(x)=1/(e^x+1)$ representing the
energy dispersion of electrons and fermi distribution function respectively.

There are, altogether, eight variables which need to be determined:
the cell velocity ${\bf v}
({\bf R})$, the cell electron temperature $T_e({\bf R})$,
the particle flux $\langle{\bf J}\rangle$, the energy flux
$\langle{\bf J}_H\rangle$, the average local kinetic energy density
$u({\bf R})$, the local number density
of electrons $n({\bf R})$, the local chemical potential $\mu({\bf R})$,
and the total electrical potential $\phi({\bf R})$. Moreover, there are
three balance equations (\ref{eq1}), (\ref{eq2}), and (\ref{eq3}),
supplemented by four relations (\ref{jav}), (\ref{jhl}),
(\ref{u}) and (\ref{n}), as well as the Poisson
equation relating electron density with potential:
\begin{equation}
\nabla^2\phi=-4\pi e[n({\bf R})-n^+]
\end{equation}
with $n^+$ as the density of the ionized donor background. These eight
equations form a close set of equations for the hydrodynamic device
modeling.

\section{Onsager relation in hydrodynamic balance equation approach}

In this section, we demonstrate the Onsager relation,\cite{onsager,mahan}
more accurately, we verify the validity of hydrodynamic balance equations
in regard to the Onsager relation. It is well known that the Onsager
relation is a manifestation of microscopic irreversibility for any
statistical system near thermal equilibrium. Therefore any properly
formulated statistical physics model should satisfy this relation. It is
very easy to verify this relation in the framework of Kubo linear response
theory. Moreover, if one can determine the distribution function from
the Boltzmann equation, it is also straightforward to verify the Onsager
relation by calculating the pertinent moments of the distribution function.
However, for the traditional hydrodynamic
model,\cite{1,2,3,4,5,6,7,8,9,10,11,12} verification has been elusive.
In fact, in a very recent article, Anile {\em et al.}\cite{anile3} showed
that the Onsager relation breaks down in this model.
Although they tried to
circumvent this difficulty, they did not establish the existence of the
relation they employed within the model itself by Monte Carlo simulation.
Here, we will examine the Onsager relation within the framework of
the hydrodynamic balance equations.

The Onsager relation\cite{mahan} is concerned with the linear response of
 the particle current $\langle{\bf J}\rangle$ and the heat
flux $\langle {\bf J}_Q\rangle$ near thermal equilibrium, which
flow as a result of forces ${\bf X}_i$ on the system:
\begin{eqnarray}
\label{onsager1}
\langle{\bf J}\rangle&=&L^{11}{\bf X}_1+L^{12}{\bf X}_2\;,\\
\label{onsager2}
\langle{\bf J}_Q\rangle&=&L^{21}{\bf X}_1+L^{22}{\bf X}_2\;,
\end{eqnarray}
with ${\bf X}_1=-\frac{1}{T}\nabla (\mu+e\phi)$ and ${\bf X}_2=\nabla(1/T)$.
The Onsager relation states that
\begin{equation}
L^{12}=L^{21}\;.
\end{equation}
The heat flux $\langle{\bf J}_Q\rangle$ relates to the energy flux in
Eq.\ (\ref{jhl}) through
\begin{equation}
\langle{\bf J}_Q\rangle=\langle{\bf J}_H\rangle-\mu\langle{\bf J}\rangle\;.
\end{equation}
The fluxes $\langle{\bf J}\rangle$ and $\langle{\bf J}_H\rangle$ have already
been defined in the previous section by Eqs.\ (\ref{jav}) and (\ref{jhl}).
Our first task is to express them in terms of linear response in the form of
Eqs.\ (\ref{onsager1}) and (\ref{onsager2}).

The first relation can be acquired directly by linearization of force
balance equation, Eq.\ (\ref{eq2}), near thermal equilibrium, so that
we only need to consider a steady state with the external electric field
${\bf E}$ and the spatial gradient being very small. Then $T_e=T$ and
${\bf v}$ is also very small. We take ${\bf E}$, $\nabla T$ and ${\bf v}$
to be in the $x$-direction and treat Eq.\ (\ref{eq2}) to first order
in the small quantities. This
means, for instance, the gradient operator $\nabla_x\equiv\partial/\partial
x$ is a first order small quantity and $v_x$ is also a first order small
quantity, thus $\nabla_x v_x$ is a higher-order small quantity and can be
neglected. These facts should be took in mind in all of our following
calculations. Therefore the force balance equation Eq.\ (eq2) can be
written as
\begin{equation}
0=-\frac{2}{3nm}\nabla_x u+\frac{eE_x}{m}+\frac{f_x}{nm}\;.
\end{equation}
All the quantities in the other two directions are zero. For small $v_x$,
$f_x$ is proportional to $v_x$,\cite{leihoring} thus proportional to
$\langle{J_x}\rangle$, and
\begin{equation}
\rho=-\frac{f_x}{n^2e^2v_x}=-\frac{f_x}{ne^2\langle J_x\rangle}\;,
\end{equation}
is the resistivity and independent of $v_x$ ($\langle J_x\rangle$), which
is given by
\begin{eqnarray}
\rho&=&-\frac{4\pi}{n^2e^2}\sum_{{\bf q}\lambda}q_x^2|M({\bf q},\lambda)|^2
\left[-\frac{1}{T}n^\prime(\frac{\Omega_{{\bf q}\lambda}}{T})\right]
\left[f(\frac{\varepsilon_{\bf k}-\mu}{T})-
f(\frac{\varepsilon_{{\bf k}+{\bf q}}-\mu}{T})\right]
\delta(\varepsilon_{{\bf k}+{\bf q}}
-\varepsilon_{\bf k}+\Omega_{{\bf q}\lambda})\nonumber\\
\label{rho}
&&\mbox{}-\frac{n_i}{n^2e^2}\sum_{\bf q}q_x^2|u({\bf q})|^2\frac{\partial}
{\partial \omega}\Pi_2({\bf q},\omega)|_{\omega=0}\;.
\end{eqnarray}
We then have
\begin{equation}
\label{j1}
\langle J_x\rangle=\frac{E_x}{e\rho}-\frac{2}{3}\frac{\nabla_xu}{ne^2\rho}\;.
\end{equation}
Employing Eqs.\ (\ref{u}) and (\ref{n}), we can express Eq.\ (\ref{j1}) in
the form of Eq.\ (\ref{onsager1}), with
\begin{eqnarray}
L^{11}&=&\frac{T}{\rho e^2}\;,\\
\label{l12}
L^{12}&=&\frac{T^2}{\rho e^2}\left[\frac{5}{3}\frac{F_{3/2}(\zeta)}
{F_{1/2}(\zeta)}\zeta\right]\;.
\end{eqnarray}
Here $\zeta=\mu /T$ and the function $F_{\nu}(y)$ is defined by
\begin{equation}
F_\nu(y)=\int_0^\infty\frac{x^\nu dx}{\exp(x-y)+1}\;.
\end{equation}

The procedure for identifying the linearized heat flux is, of course,
similar to that of particle flux. Therefore we consider the
rate of change of the energy flux operator ${\bf J}_H$ defined by
Eq.\ (\ref{jhoper}):
\begin{eqnarray}
\dot {\bf J}_H({\bf R})&=&-i[{\bf J}_H({\bf R}),H]=-\nabla\cdot {\cal A}
\nonumber\\
&&\mbox{}+\frac{1}{2}\sum_i\frac{(e{\bf E}+{\bf F}_i)\cdot{\bf p}_i}
{2m}\frac{{\bf
p}_i}{m}\delta({\bf r}_i-{\bf R})+\frac{1}{2}\sum_i
\frac{{\bf p}_i\cdot(e{\bf E}+{\bf F}_i)}{2m}\frac{{\bf
p}_i}{m}\delta({\bf r}_i-{\bf R})\nonumber\\
&&\mbox{}+\frac{1}{2}\sum_i\frac{{\bf p}_i^2}{2m}\frac{e{\bf E}+{\bf F}_i}
{m}\delta({\bf r}_i-{\bf R})+\frac{1}{2}\sum_i\frac{e{\bf E}+{\bf F}_i}
{m}\frac{{\bf p}_i^2}{2m}\delta({\bf r}_i-{\bf R})\nonumber\\
\label{jhdot}
&&\mbox{}+\frac{1}{2}\sum_i\frac{{\bf p}_i}{m}\frac{(e{\bf E}+{\bf F}_i)
\cdot{\bf p}_i}{2m}\delta({\bf r}_i-{\bf R})+\frac{1}{2}\sum_i\frac{{\bf
p}_i}{m}\frac{{\bf p}_i\cdot(e{\bf E}+{\bf F}_i)}{2m}
\delta({\bf r}_i-{\bf R})\;,
\end{eqnarray}
where we have used Eqs.\ (\ref{rdot}) and (\ref{pdot}) again. The tensor
${\cal A}$ is defined as
\begin{equation}
\label{a}
{\cal A}=\sum_i\frac{{\bf p}_i^2}{2m}\frac{{\bf p}_i}{m}\frac{{\bf p}_i}{m}
\delta({\bf r}_i-{\bf R})\;.
\end{equation}
Performing the statistical average of Eq.\ (\ref{jhdot}), we have
\begin{equation}
\label{eq4}
\langle\dot{\bf J}_H\rangle+\nabla\cdot\langle{\cal A}\rangle
=\langle{\bf B}\rangle+\frac{5}{3m}eu{\bf E}+en{\bf E}\cdot{\bf v}{\bf v}
+\frac{1}{2}env^2{\bf E}+\frac{1}{2}v^2{\bf f}-w{\bf v}\;.
\end{equation}
It is understood that the right hand side of Eq.\ (\ref{eq4}) is derived by
transforming the coordinate and moment operators to the relative variables
of Eq.\ (\ref{pprp}), before performing the statistical averages.
The expression of $\langle {\bf B}\rangle$ is given in the Appendix,
and $\langle{\cal A}\rangle$ can be expressed as
\begin{equation}
\label{aa}
\langle{\cal A}\rangle=\frac{1}{3}(S({\bf R})+uv^2){\cal I}+\langle{\bf J}
_H\rangle{\bf v}+{\bf v}\langle{\bf J}_H\rangle-u{\bf v}{\bf v}-\frac{1}{2}
mnv^2{\bf v}{\bf v}\;,
\end{equation}
with
\begin{equation}
S({\bf R})=\left\langle\sum_i\frac{{\bf p}_i^{\prime 4}}{2m^3}
\delta({\bf r}_i^\prime)\right\rangle\;.
\end{equation}
This average can be calculated in the balance equation theory mentioned
using the density matrix ${\hat\rho}$ discussed in the previous section,
with the result
\begin{equation}
\label{sr}
S({\bf R})=2\sum_{\bf k}\frac{k^4}{2m^3}f(\frac{\varepsilon_{\bf k}-\mu}
{T_e})\;.
\end{equation}

It should be emphasized here that if the density matrix employed in the
balance equation is exactly the real physical one, then Eq.\ (\ref{eq4})
should be consistent with Eqs.\ (\ref{eq1})-(\ref{eq3}). This is to say
that if we have calculated every unknown parameters from the hydrodynamic
balance equations presented in the previous section, and substitute them in
Eq.\ (\ref{eq4}), then Eq.\ (\ref{eq4}) should merely be an identity.
Unfortunately,
in actual fact, this is not the case, especially when the system is a bit
far away from weakly nonuniform system. However, here we do not care about
it, because we only need this equation holds near thermal equilibrium. In
this circumstance, the density matrix, chosen in balance equation theory,
has already been shown to be reasonable, in particular for a system with
strong electron-electron interactions.\cite{chen1,chen2} Therefore
Eq.\ (\ref{eq4}) should yield agreement with the balance equations near
thermal equilibrium, and we may use it to determine the linear response
relation of $\langle{\bf J}_H\rangle$ with the external forces ${\bf X}_i$
and examine whether the result obtained satisfies Onsager relation.

Thus, to the first order in the small quantities, Eq.\ (\ref{eq4}) can
be written in the form
\begin{equation}
\label{j20}
\frac{5}{3m}eu({\bf R})E_x-\frac{1}{3}\nabla_x S({\bf R})+\langle B_x
\rangle=0\;.
\end{equation}
In deriving this equation, we have used the linearized force and energy
balance equations, Eqs.\ (\ref{eq2lei}) and (\ref{eq3lei}),
and $\langle B_x\rangle$ has also been linearized and is proportional
to $\langle{\bf J}_H\rangle$,
which is $\frac{5}{3}uv_x$ to first order. Thus we may define
\begin{equation}
\frac{1}{\tau}=\frac{\langle{\bf B}\rangle}{n({\bf R})
\langle{\bf J}_H\rangle}\;,
\end{equation}
which is also independent of $v_x$ ($\langle{\bf J}_H\rangle$).
Substituting this relation into Eq.\ (\ref{j20}) and calculating the gradient
of $S({\bf R})$ in Eq.\ (\ref{sr}), we find the average energy flux is
given by
\begin{equation}
\langle{\bf J}_H\rangle=-\frac{5}{3}\frac{T^2}{m}\frac{F_{3/2}(\zeta)}
{F_{1/2}(\zeta)}\tau{\bf X}_1-\frac{T^3}{m}
\left[\frac{7}{3}\frac{F_{5/2}(\zeta)}
{F_{1/2}(\zeta)}{F_{1/2}(\zeta)}-\frac{5}{3}\zeta\frac{F_{3/2}(\zeta)}{F_{1/2}
(\zeta)}\right]\tau{\bf X}_2\;.
\end{equation}
Subtracting $\mu\langle{\bf J}\rangle$,
we obtain the linearized heat flux in terms of ${\bf X}_1$ and ${\bf X}_2$
and can identify the linear coefficients of Eq.\ (\ref{onsager2}) as
\begin{eqnarray}
\label{l21}
L^{21}&=&\frac{T^2}{\rho e^2}\left[-\frac{\tau\rho e^2}{m}\frac{5}{3}
\frac{F_{3/2}(\zeta)}{F_{1/2}(\zeta)}-\zeta\right]\;,\\
L^{22}&=&-\frac{\tau T^3}{m}\left[\frac{7}{3}\frac{F_{5/2}(\zeta)}
{F_{1/2}(\zeta)}{F_{1/2}(\zeta)}-\frac{5}{3}\zeta\frac{F_{3/2}(\zeta)}{F_{1/2}
(\zeta)}\right]-\frac{\zeta T^3}{\rho e^2}\left[\frac{5}{3}
\frac{F_{3/2}(\zeta)}{F_{1/2}(\zeta)}-\zeta\right]\;.
\end{eqnarray}
Comparing Eq.\ (\ref{l21}) with Eq.\ (\ref{l12}), we find that the condition
under which the Onsager relation holds is given by
\begin{equation}
\label{id}
I\equiv-\frac{\tau\rho e^2}{m}=1\;.
\end{equation}

We have closely examined Eq.\ (\ref{id}) for a GaAs system to see if it is
indeed satisfied in balance equation theory. Both $\rho$ (Eq.\ (\ref{rho}))
and $\langle B_x\rangle$ (Appendix) are composed of contributions due to
electron-impurity, electron--LO-phonon, and electron--Ac-phonon scatterings
(with the electron--acoustic-phonon scatterings due to longitudinal mode
acoustic phonons via deformation potential and piezoelectric
interactions, and transverse mode via piezoelectric interaction).
We have examined each scattering contribution in detail to check
Eq.\ (\ref{id}) separately for each interaction. It is clear that if
$-\frac{e^2\rho_i/m}{(1/\tau)_i}=1$ is satisfied for each interaction,
we have $-\frac{e\sum_i\rho_i/m}{\sum_i(1/\tau)_i}=1$.
Moreover, this procedure effects the fact that the result should be
independent of impurity concentration and parameters of the electron-phonon
interaction matrixes.

The expressions for $I$ obtained from the balance equations are given by
\begin{equation}
\label{iei}
I_{ei}=\frac{\sum_{\bf q}q^2|u({\bf q})|^2[\frac{\partial}{\partial \omega}
\Pi_2^\varepsilon({\bf q},\omega)]|_{\omega=0}}{(\frac{5}{3})(\frac{u}{n})
\sum_{\bf q}q^2|u({\bf q})|^2[\frac{\partial}{\partial \omega}
\Pi_2({\bf q},\omega)]|_{\omega=0}}\;,
\end{equation}
due to electron-impurity scattering; and
\begin{eqnarray}
I_{e-ph}(\lambda)&=&\frac{\sum_{\bf q}|M({\bf q},\lambda)|^2\Omega_{{\bf q}
\lambda}(\varepsilon_{\bf q}+\Omega_{{\bf q}\lambda})n^\prime(\frac{\Omega_
{{\bf q}\lambda}}{T})\Pi_2({\bf q},\Omega_{{\bf q}\lambda})}{(\frac{5}{3})
(\frac{u}{n})\sum_{\bf q}|M({\bf q},\lambda)|^2\frac{{\bf q}^2}{m}
n^\prime(\frac{\Omega_{{\bf q}\lambda}}{T})\Pi_2({\bf q},
\Omega_{{\bf q}\lambda})}\nonumber\\
\label{ieph}
&&\mbox{}+\frac{-\sum_{\bf q}|M({\bf q},\lambda)|^2\frac{{\bf q}^2}{m}
n^\prime(\frac{\Omega_{{\bf q}\lambda}}{T})\Pi_2^\varepsilon({\bf q},
-\Omega_{{\bf q}\lambda})}{(\frac{5}{3})
(\frac{u}{n})\sum_{\bf q}|M({\bf q},\lambda)|^2\frac{{\bf q}^2}{m}
n^\prime(\frac{\Omega_{{\bf q}\lambda}}{T})\Pi_2({\bf q},
\Omega_{{\bf q}\lambda})}\;,
\end{eqnarray}
due to electron-phonon scattering, for phonons of mode $\lambda$.
$I_{e-ph}(\lambda)$ is further composed of contributions due to
electron--LO-phonon scattering, $I_{e-LO}$; due to  electron--longitudinal
acoustic phonons by deformation potential coupling, $I_{edl}$;
and by piezoelectric interaction, $I_{epl}$; and due to
electron--transverse acoustic phonons by piezoelectric
interaction, $I_{ept}$. $\Pi_2^\varepsilon$ in Eqs.\ (\ref{iei})
and (\ref{ieph}) is defined by
\begin{equation}
\Pi_2^\varepsilon({\bf q},\omega)=2\pi\sum_{\bf k}\varepsilon_{\bf k}
\delta(\varepsilon_{{\bf k}+{\bf q}}-\varepsilon_{\bf k}+\omega)
\left[f\left(\frac{\varepsilon_{\bf k}-\mu}{T}\right)-
f\left(\frac{\varepsilon_{{\bf k}+{\bf q}}-\mu}{T}\right)\right]\;.
\end{equation}
For the LO phonon, $\Omega_{{\bf q},LO}=\Omega_0=35.4$\ meV, and the
Fr\"olich matrix element is $|M({\bf q},LO)|^2=e^2(\kappa_\infty^{-1}
-\kappa^{-1})\Omega_0/(2\varepsilon_0q^2)\propto 1/q^2$. (Since the
constants in the matrix elements cancel in Eq.\ (\ref{ieph}), therefore
in the following we only specify their relation to $q$.) The matrix
element due to longitudinal deformation potential coupling is
$|M({\bf q},dl)|^2\propto q$, that due to longitudinal
piezoelectric interaction is $|M({\bf q},pl)|^2\propto(q_xq_yq_z)^2/q^7$,
and for the two branches of independent transverse piezoelectric
interaction: $\sum_{j=1,2}|M({\bf q},pt_j)|^2\propto (q_x^2q_y^2+q_y^2q_z^2
+q_z^2q_x^2-(3q_xq_yq_z)^2/q^2)/q^5$. For acoustic phonons $\Omega_{{\bf q}
\lambda}$ can be written as $v_{s}q$, with the longitudinal sound speed
$v_s$ being 5.29$\times 10^3$\ m/s, and the transverse sound speed being
2.48$\times 10^3$\ m/s. The effective mass of electron is $0.07m_e$,
with $m_e$ denoting the free electron mass.

The results of  our numerical calculations are presented in Fig.\ 1 to
Fig.\ 5, where contributions to  $I$ due to the various interactions
discussed above are plotted against electron density for several different
temperatures. As it is generally believed that the contribution of acoustic
phonons is important only at low temperature, while the contribution
of LO phonons is dominant at high temperature, our temperatures are chosen
as 10, 20, and 40\ K for the former, and 50, 300, 500, and 1000\ K for
the letter. Impurity scattering is present at any temperature,
so we take $T=$10, 50, 100, 300, and 1000\ K in Fig.\ 1. From these
figures it is evident that, for any temperature, when electron density is
sufficiently high $I$ is exactly unity, indicating that the Onsager relation
holds. It is also seen from the figures as temperature becomes higher,
the electron density needed to make the Onsager relation hold is also higher.
An interesting exception is the LO phonon in Fig.\ 2, in which we can see
that the needed density for $T=300$\ K is lower than that for $T=50$\ K,
to assure that $I_{eLO}=1$.

\section{Conclusions and Discussions}

In this paper, we have clarified the role of heat flux in hydrodynamic
balance equations. We have further shown that, for any temperature, when
electron density is sufficiently high, the hydrodynamic balance equation
theory satisfies the Onsager relation. This is consistent with the
understanding that the Lei-Ting balance equation theory holds only for
strong electron-electron interactions. Our result supports the validity of
this theory in a weakly nonuniform system. To our knowledge, this is the
first set of hydrodynamic equations which satisfies the Onsager relation
self contained and without the {\em ad hoc} introduction of terms which
do not originate within the theory.

However, we should also point out that the hydrodynamic balance equations can
only be used to describe weakly nonuniform systems. When the temperature
gradient is large, and/or there is a large heat flux in the system, for
example in phenomena as impact ionization and heat generation in
nonuniform systems, the energy flux equation (Eq.\ (\ref{eq4})), or heat
flux equation, which is of paramount importance in describing these
phenomena, is no longer consistent with the other balance equations
(Eqs.\ (\ref{eq1})-(\ref{eq3})), and a contradiction emerges. This
reflects the inadequacy of the assumed initial density matrix,
Eq.\ (\ref{rho0}), in Lei-Ting balance equation theory, by failing to
include the detailed information about the physics of heat flux. This can be
further illustrated as follows: In our deriving the average energy flux
operator Eq.\ (\ref{jhoper}), there should be another term
\begin{equation}
\langle{\bf j}_H\rangle=\langle\sum_i\frac{{\bf p}_i^{\prime 2}}{2m}
\frac{{\bf p}_i^\prime}{m}\delta({\bf r}_i^\prime)\rangle
\end{equation}
on the right hand side of Eq.\ (\ref{jhl}). Moreover, in obtaining the
average of the tensor ${\cal A}$ in Eq.\ (\ref{a}), there should be another
term ${\bf v}\cdot\langle\sum_i\frac{{\bf p}_i^\prime}{m}
\frac{{\bf p}_i^\prime}
{m}\frac{{\bf p}_i^\prime}{m}\delta({\bf r}_i^\prime)\rangle$ on the right
hand side of Eq.\ (\ref{aa}). These two terms do not vanish when the system
is not near thermal equilibrium, and should be included in the theory
if they are calculated from a {\em real} physical density matrix. Anile
{\em et al}.\cite{anile3} have included such terms in their traditional
hydrodynamic equations mentioned in the introduction. Unfortunately these
terms are exactly zero predicted by balance equation theory.

It is clear that for mediately nonuniform systems and/or systems
far from thermal equilibrium, an accurate prediction of the behavior of
heat flux requires the inclusion of one or more additional unknown
parameters in the initial density matrix (in high-order terms so that
they do not violate the particle and momentum balance equations) to be
followed by their determination from expanded balance equations, which now
include the heat flux equation(s).
This problem is currently under investigation, and the results
will be published in elsewhere.

\acknowledgements

One of the authors (MWW) would like to thank Professor X.L. Lei, who first
brought this problem into his attention. This research is supported by U.S.
Office Naval Research (Contract No. N66001-95-M-3472), and the U.S. Army
Research Office.

\appendix
\section*{}

The expression of $\langle{\bf B}\rangle$ is composed of two parts. One is
due to collisions with impurities ($\langle{\bf B}_i\rangle$), and the other
is due to interaction with phonons ($\langle{\bf B}_{ph}\rangle$). They
are given by
\begin{eqnarray}
\langle{\bf B}_i\rangle&=&2\pi n_i\sum_{\bf kq}|u({\bf q})|^2(\varepsilon_{
{\bf k}+{\bf q}}-\varepsilon_{\bf k})\frac{{\bf k}+{\bf q}/2}{m}\delta(
\varepsilon_{{\bf k}+{\bf q}}-\varepsilon_{\bf k}+{\bf q}\cdot{\bf v})
\nonumber\\
&&\hspace{1cm}\mbox{}\times\left[f(\frac{\varepsilon_{\bf k}-\mu}{T_e})-
f(\frac{\varepsilon_{{\bf k}+{\bf q}}-\mu}{T_e})\right]\nonumber\\
&&\mbox{}+2\pi n_i\sum_{\bf kq}|u({\bf q})|^2({\bf q}\cdot{\bf v}\frac
{{\bf k}+{\bf q}}{m}+{\bf k}\cdot{\bf v}\frac{{\bf q}}{m})\delta(\varepsilon
_{{\bf k}+{\bf q}}-\varepsilon_{\bf k}+{\bf q}\cdot{\bf v})\nonumber\\
&&\hspace{1cm}\mbox{}\times\left[f(\frac{\varepsilon_{\bf k}-\mu}{T_e})-
f(\frac{\varepsilon_{{\bf k}+{\bf q}}-\mu}{T_e})\right]\nonumber\\
&&\mbox{}+\pi n_i\sum_{\bf kq}|u({\bf q})|^2(\varepsilon_{
{\bf k}+{\bf q}}+\varepsilon_{\bf k})\frac{{\bf q}}{m}\delta(
\varepsilon_{{\bf k}+{\bf q}}-\varepsilon_{\bf k}+{\bf q}\cdot{\bf v})
\nonumber\\
&&\hspace{1cm}\mbox{}\times\left[f(\frac{\varepsilon_{\bf k}-\mu}{T_e})-
f(\frac{\varepsilon_{{\bf k}+{\bf q}}-\mu}{T_e})\right]\;,
\end{eqnarray}
and
\begin{eqnarray}
\langle{\bf B}_{ph}
\rangle&=&-4\pi\sum_{{\bf kq}\lambda}|M({\bf q},\lambda)|^2
(\varepsilon_{{\bf k}+{\bf q}}-\varepsilon_{\bf k})
\frac{{\bf k}+{\bf q}/2}{m}\delta(\varepsilon_{{\bf k}+{\bf q}}-
\varepsilon_{\bf k}+\Omega_{{\bf q}\lambda}-{\bf q}\cdot{\bf v})
\nonumber\\
&&\hspace{1cm}\mbox{}\times\left[f(\frac{\varepsilon_{\bf k}-\mu}{T_e})-
f(\frac{\varepsilon_{{\bf k}+{\bf q}}-\mu}{T_e})\right]\left[n(\frac{\Omega_
{{\bf q}\lambda}}{T})-n(\frac{\Omega_{{\bf q}\lambda}-{\bf q}\cdot{\bf v}}
{T_e})\right]\nonumber\\
&&\mbox{}-4\pi\sum_{{\bf kq}\lambda}|M({\bf q},\lambda)|^2
({\bf q}\cdot{\bf v}\frac{{\bf k}+{\bf q}}{m}+{\bf k}\cdot{\bf v}
\frac{{\bf q}}{m})\delta(\varepsilon_{{\bf k}+{\bf q}}-
\varepsilon_{\bf k}+\Omega_{{\bf q}\lambda}-{\bf q}\cdot{\bf v})
\nonumber\\
&&\hspace{1cm}\mbox{}\times\left[f(\frac{\varepsilon_{\bf k}-\mu}{T_e})-
f(\frac{\varepsilon_{{\bf k}+{\bf q}}-\mu}{T_e})\right]\left[n(\frac{\Omega_
{{\bf q}\lambda}}{T})-n(\frac{\Omega_{{\bf q}\lambda}-{\bf q}\cdot{\bf v}}
{T_e})\right]\nonumber\\
&&\mbox{}-2\pi\sum_{{\bf kq}\lambda}|M({\bf q},\lambda)|^2
(\varepsilon_{{\bf k}+{\bf q}}+\varepsilon_{\bf k})
\frac{{\bf q}}{m}\delta(\varepsilon_{{\bf k}+{\bf q}}-
\varepsilon_{\bf k}+\Omega_{{\bf q}\lambda}-{\bf q}\cdot{\bf v})
\nonumber\\
&&\hspace{1cm}\mbox{}\times\left[f(\frac{\varepsilon_{\bf k}-\mu}{T_e})-
f(\frac{\varepsilon_{{\bf k}+{\bf q}}-\mu}{T_e})\right]\left[n(\frac{\Omega_
{{\bf q}\lambda}}{T})-n(\frac{\Omega_{{\bf q}\lambda}-{\bf q}\cdot{\bf v}}
{T_e})\right]\;.
\end{eqnarray}

\begin{figure}
\caption{$I$ due to electron-impurity scattering is plotted as a function of
electron density for several different temperatures}
\end{figure}

\begin{figure}
\caption{$I$ due to electron--LO-phonon scattering is plotted
as a function of electron density for several different temperatures}
\end{figure}

\begin{figure}
\caption{$I$ due to electron--longitudinal acoustic-phonon scattering via
deformation potential coupling is plotted
as a function of electron density for several different temperatures}
\end{figure}

\begin{figure}
\caption{$I$ due to electron--longitudinal acoustic-phonon scattering via
piezoelectric interaction is plotted
as a function of electron density for several different temperatures}
\end{figure}

\begin{figure}
\caption{$I$ due to electron--transverse acoustic-phonon scattering via
piezoelectric interaction is plotted
as a function of electron density for several different temperatures}
\end{figure}


\begin{references}

\bibitem{leiting} X.L. Lei and C.S. Ting, Phys. Rev. B {\bf
30}, 4809 (1984); {\bf 32}, 1112 (1985).
\bibitem{leihoring} X.L. Lei and N.J.M. Horing, Int. J. Mod. Phys. B {\bf 6},
805 (1992).
\bibitem{hirakawa} K. Hirakawa and H. Sakaki, J. Appl. Phys. {\bf 63},
803 (1988).
\bibitem{lei} X.L. Lei, J. Cai, and L.M. Xie, Phys. Rev. B {\bf 38}, 1529
(1988).
\bibitem{1} Q. Lin, N. Goldsman, and G. Tai, Solid-State Electron. {\bf 37},
359 (1994).
\bibitem{2} S. Ramaswamy and T. Tang, IEEE Trans. Ele. Dev. {\bf ED-41}, 76
(1994)
\bibitem{3} G.L. Gardner, SIAM J. Appl. Math. {\bf 54}, 409 (1994).
\bibitem{4} Y. Zhang and M. Nokali, Solid-State Electron. {\bf 36}, 1689
(1993).
\bibitem{5} J.R. Zhou, D. Vasileska, and D.K. Ferry, Solid-State
Electron. {\bf 36}, 1294 (1993).
\bibitem{6} T. Tang, S. Ramaswamy, and J. Nam, IEEE Trans. Ele.
Dev. {\bf 40}, 1469 (1993).
\bibitem{7} A. Majorana and G. Russo, COMPEL-Int. J. Comp. Math. Electrical
and Electron. Eng. {\bf 12}, 81 (1993).
\bibitem{8} D.K. Ferry and J.R. Zhou, Phys. Rev. B {\bf 48}, 7944 (1993).
\bibitem{9} D.A. Teeter, J.R. East, RIK. Mains, and G.I. Haddad, IEEE
Trans. Ele. Dev. {\bf ED-40}, 837 (1993).
\bibitem{10} J.R. Zhou and D.K. Ferry, IEEE Trans. Ele. Dev. {\bf ED-40},
421 (1993).
\bibitem{11} K. Souissi, F. Odeh, H.H.K. Tang, A. Gnudi, and P.F. Lu,
IEEE Trans. Ele. Dev. {\bf ED-40}, 1501 (1993).
\bibitem{12} W. Quade, E. Sch\"oll, and M. Rudan, Solid-State Electron.
{\bf 36}, 1493 (1993).
\bibitem{anile1} A.M. Anile and S. Pennisi, Phys. Rev. B {\bf 46},
13186 (1992).
\bibitem{anile2} A.M. Anile and S. Pennisi, Continuum Mech. Thermodyn.
{\bf 4}, 187 (1992).
\bibitem{anile3} A.M. Anile and O. Muscato, Phys. Rev. B {\bf 51},
16728 (1995).
\bibitem{chen1} L.Y. Chen, C.S. Ting, and N.J.M. Horing, Phys. Rev. B {\bf
42}, 1129 (1990).
\bibitem{chen2} L.Y. Chen, C.S. Ting, and N.J.M. Horing, Solid State Commun.
{\bf 73}, 437 (1990).
\bibitem{cai1} J. Cai, H.L. Cui, and N.J.M. Horing, and X.L. Lei, Mat.
Res. Soc. Symp. Proc. {\bf 326}, 215 (1994).
\bibitem{cai2} J. Cai, H.L. Cui, N.J.M. Horing, X.L. Lei, E. Lenzing, and
B.S. Perlman, Proc. 3rd Int. Workshop Comp. Electron., p.\ 127, 1994.
\bibitem{cai3} J. Cai and H.L. Cui, J. Appl. Phys. (1995).
\bibitem{onsager} L. Onsager, Phys. Rev. B {\bf 37}, 405 (1931).
\bibitem{mahan} G.D. Mahan, {\it Many-Particle Physics}, (Plenum Press,
New York 1981), p.\ 212.
\bibitem{scalapino} D.J. Scalapino, in {\it Superconductivity}, ed. by R.D.
Parks, Vol.\ 1, (Marcel Dekker, New York), 1969, p449.
\bibitem{callaway} J. Callaway, {\it Quantum Theory of the Solid State},
2nd. ed., (San Diego, Academic, 1991).
\bibitem{zubarev} D.N. Zubarev, {\it Nonequilibrium Statistical
Thermodynamics}, (New York, Consultants Bureau, 1974).

\end{references}
\end{document}